\documentclass[floatfix,showpacs,12pt,nofootinbib]{revtex4}
\usepackage{amssymb,amsmath}
\usepackage{graphicx,float}
\usepackage{indentfirst}

\newcommand{\beq}{\begin{equation}}
\newcommand{\eeq}{\end{equation}}
\newcommand{\bea}{\begin{eqnarray}}
\newcommand{\eea}{\end{eqnarray}}

\DeclareMathOperator{\sech}{sech}
\def\({\left(}
\def\){\right)}

\begin{document}

\title{A method for obtaining thick brane models}
\author{A. de Souza Dutra} \email{dutra@feg.unesp.br}
\author{G. P. de Brito} \email{gustavopazzini@gmail.com}
\author{J. M. Hoff da Silva} \email{hoff@feg.unesp.br}
\affiliation{Departamento de F\'{\i}sica e Qu\'{\i}mica, Universidade Estadual Paulista,
Av. Dr. Ariberto Pereira da Cunha, 333, Guaratinguet\'a, SP, Brazil.}

\begin{abstract}
In this work we present a new class of braneworld models, where the superpotential function may be decomposed in a very specific form, but still embracing, which allows considerable simplifications in the field equations. We exemplify the method with two specific models that are extended versions of superpotentials already considered in the literature. As one can see in the examples, the braneworld scenarios obtained here present very interesting features, such as split brane mechanism and asymmetric warp factor shape, which is essential to address the hierarchy problem in the context of thick brane scenarios.
\end{abstract}

\maketitle
\section{Introduction}

In the last decade the idea that our four-dimensional Universe is embedded in a higher dimensional spacetime with warped geometry has received considerable amount of attention in the literature (for some comprehensive reviews see refs. \cite{Csaki}-\cite{Rubakov}). In fact the very first motivation for the introduction of braneworld models constructed upon warped spacetime was the solution of the hierarchy problem, as realized by L. Randall and R. Sundrum \cite{RS1}. Another interesting feature regarding warped geometry is that it is possible to construct models containing noncompact extra dimensions, without loosing compatibility with the gravitational experiments \cite{RS2}.

Since then, several extensions of the RS model have been proposed generalizing it in a broad range of aspects. A very interesting research program was the one initiated by De Wolfe \textit{et. al.} \cite{DEWOLFE} and, independently, by Gremm \cite{GREMM}, where the full spacetime is a five-dimensional manifold with warped geometry and, our four-dimensional Universe is interpreted to be a thick brane generated by a scalar field coupled to gravity (see \cite{folomeev} for an updated review in thick branes scenarios). In this context, the brane is interpreted as being a domain-wall where the standard model fields are localized in.

The developing of thick brane models has been achieving many important results \cite{Melfo}-\cite{Bazeia}. We emphasize the possibility of splitting brane mechanism, where the resulting setup supports two different branes corresponding to a double kink solution in the scalar sector \cite{amaro}. Recently, it was also proposed that the splitting brane mechanism, along with asymmetric properties in the warp factor, may be used to address the hierarchy problem in the context of thick brane scenarios \cite{ahmed,hoff}. 

Since those scenarios containing two different thick branes arise as a consequence of double kinks, it is natural to ask about the possible construction of multi-brane configurations related with the so-called multikink solutions. In fact, multikink configurations may be useful in modeling physical systems with many domain walls. In this vein, Peyrard and Kruskal \cite{kruskal} discovered that a single kink becomes unstable when it moves in a discrete lattice with large velocities, while multikink solutions remain stable. This effect is associated with the interaction between the kink and the radiation, and the resonances were already experimentally observed \cite{malomed,orlando}. However, some few works addressed the problem of obtaining analytical models presenting multikink solutions \cite{brito1,brito2}.

In this work, it is our intention to implement an approach which allows one to achieve both asymmetric and multi-wall branes. As we will see, multi-brane scenarios are not necessarily related with multikinks and, we will construct multi-brane configurations arising from scalar field models supporting usual kink solutions. The asymmetric resulting warp factor models are certainly appealing for approaching the hierarchy problem, while the warp factor symmetric multi-wall generalizations are interesting for braneworld modeling. 

This paper is organized as follows: in the next section we present the basic main steps of the method whose outcomes are obtained from the superpotential choices. In Section III we show how the polynomial superpotential can lead to asymmetric as well as multi-wall solutions to the background. Section IV is devoted to some extensions of the paradigmatic Gremm's model. Basically, it is shown how to get asymmetric and multi-wall solutions as well, by applying the simple aforementioned method. In Section V we depict the stability issue concerning metric fluctuation in general grounds and in the last section we conclude.   

\section{The Framework}

We consider a braneworld scenario, in $4+1$ dimensions, described by the Einstein-Hilbert action coupled to a system of $n$ real scalar fields, namely

\begin{equation}  \label{action}
S = \int d^5x \sqrt{-g} \bigg(\frac{1}{4}R - \frac{1}{2}\sum_{i=1}^n \partial_A\phi_i \partial^A \phi_i - V(\phi_1,...,\phi_n) \bigg),
\end{equation} 
where $g = det(g_{AB})$. We use capital Latin index to label the coordinates of the five dimensional space-time, while small Latin index are used to label the scalar fields. The coordinates in the brane are represented by $ x^{\mu}$ ($\mu = 0,1,2,3$) while the bulk coordinate is denoted $x^4 = r$. The background is a warped space-time, being the line element written as

\begin{equation}
ds^2 = g_{AB}dx^A dx^B = e^{2A(r)} \eta_{\mu\nu} dx^{\mu} dx^{\nu} + dr^2,
\end{equation} 
where $\eta_{\mu\nu}$ is the usual Minkowski metric with $diag(-,+,+,+)$, and $e^{2A(r)}$ is the so-called warp factor. We assume that the warp factor and the scalar fields only depends on the extra dimension $r$. For the above system, we obtain the following set of equations of motion

\begin{equation}  \label{secondorder1}
\frac{d^2 \phi_i}{dr^2} + 4 \frac{dA}{dr} \frac{d\phi_i}{dr} = \frac{\partial V}{\partial \phi_i}, \quad (i=1,...,n) ;
\end{equation}

\begin{equation}  \label{secondorder2}
\frac{d^2 A}{dr^2} + \frac{2}{3} \sum_{i=1}^n \bigg(\frac{d\phi_i}{dr}\bigg)^2 = 0 \quad \textmd{and} \quad \bigg(\frac{dA}{dr} \bigg)^2 = \frac{1}{6}\sum_{i=1}^n \bigg(\frac{d\phi_i}{dr}\bigg)^2 - \frac{1}{3}V(\phi_1,...,\phi_i) .
\end{equation} 

As the current approach, we consider that the potential function $V(\phi_1,...,\phi_i)$ may be written in terms of a superpotential $W(\phi_1,...,\phi_i)$ in the following way

\begin{equation}\label{potential}
V(\phi_1,...,\phi_i) = \frac{1}{2}\sum_{i=1}^n \left(\frac{\partial W}{\partial \phi_i}\right)^2 - \frac{4}{3} W^2 .
\end{equation} 

In this case, it is possible to obtain a set of first-order equations that share the same solutions with (\ref{secondorder1}) and (\ref{secondorder2}), namely

\begin{equation}  \label{firstorder}
\frac{d \phi_i}{dr} = \frac{\partial W}{\partial \phi_i} \quad , \quad \frac{dA}{dr} = -\frac{2}{3}W \quad (i=1,...,n).
\end{equation}

Notice that Eq. (\ref{secondorder1}) have $2n$ integration constants (2 for each field) whilst the first Eq. (4) has two integration constants. The second Eq. (4) can be faced as consistency equation, which after all constrains one of the constants. Therefore we are left with $2n+1$ integration constants. In general, for a given potential, Eq. (\ref{potential}) gives $1$ integration constant and, on the other hand, there are $n+1$ constants coming from Eqs. (\ref{firstorder}). Hence for one scalar field the number of integration constants in the same for both systems of equations. This is the root of the argument used on Ref. \cite{DEWOLFE}. In our case, the system comprised by Eqs. (\ref{secondorder1}) and (\ref{secondorder2}) can have more solutions than that reduced one, since $2n+1\geq n+2$ for $n\geq 1$. Nevertheless, as we shall see, the obtained solutions are quite satisfactory provided the physical consequences they have. Of course, it may be relevant to look for the other possible solutions.

At this point we will particularize the class of superpotentials that will be considered here showing that, apart considerable simplification in the calculations, this class of superpotential may generate very interesting braneworld scenarios, in particular regarding the hierarchy problem. We consider that $W(\phi_1,...,\phi_n)$ may be written as a sum of superpotentials $W_i(\phi_i)$, where $W_i$ depends only on the field $\phi_i$, \textit{i.e.},

\begin{equation}
W(\phi_1,...,\phi_n) = \sum_{i=1}^n W_i(\phi_i).
\end{equation} 
In such a case we have $\partial W_i(\phi_i)/ \partial \phi_j = 0$ for $i \neq j$, therefore the first equation in (\ref{firstorder}) become decoupled

\begin{equation}  \label{firstorder2}
\frac{d\phi_i}{dr} = \frac{\partial W_i(\phi_i)}{\partial \phi_i}, \quad (i = 1,...,n).
\end{equation} 
Upon integration we obtain

\begin{equation}
\int^{\phi_i(r)}_{\phi_i(r_i)} \frac{d \phi_i}{\partial W_i} = r - r_i, \quad (i = 1,...,n),
\end{equation} 
where we have defined $\partial W_{i}\equiv \partial W_{i}/\partial \phi _{i}$. The integration constant $r_{i}$ arises as a consequence of the translational invariance and, as will be better visualized with a concrete example, the choice of this constant will determine the position of the branes. Now, notice that it is possible to solve the second equation (\ref{firstorder}) in a very simple way. In fact, let $\{A_1(r),A_2(r),\cdots,A_{n}(r)\}$ a set of solutions, supposedly known, of the following set of equations
\begin{equation}  \label{Ai}
\frac{dA_i}{dr} = -\frac{2}{3}W_i(\phi_i), \quad (i = 1,...,n).
\end{equation} 
Then, it is easy to check that the following expression satisfies the second equation in (\ref{firstorder})
\begin{equation}
A(r) = \sum_{i=1}^n A_i(r).
\end{equation}
%
%
%
%
\indent We shall consider scalar fields models that engenders domain walls solutions. According to the usual interpretation, each scalar field is associated to a given thick brane, the domain wall itself. In order to analyze the brane formation within this framework, it will be important to investigate both the energy density and the scalar curvature associated to the models. The energy density related to the scalar field sector of (\ref{action}) is given by the following expression

\begin{equation}\label{energy}
\varepsilon(r) = e^{2A(r)} \bigg(\frac{1}{2} \sum_{i=1}^n \left(\frac{d \phi_i}{dr}\right)^2 + V(\phi_1 ,...,\phi_n) \bigg).
\end{equation} 
We may use the first-order formalism above exposed in order to rewrite the last equation as

\begin{equation}
\varepsilon(r) = -\frac{3}{4}\frac{d^2}{dr^2}(e^{2A(r)}).
\end{equation} 
On the other hand, the scalar curvature computed for the warped metric may be written in the following way

\begin{equation}
R(r) = - \bigg( 20 \left(\frac{dA}{dr}\right)^2 + 8 \frac{d^2 A}{dr^2} \bigg).
\end{equation} 
We interpreted that those regions where the branes are located, corresponds to transition regions in the scalar of curvature. Henceforth, after this general approach, let us particularize our study to relevant braneworld models. The particularization will be done by means of the number of scalar fields in Sections III and IV and, most importantly, by setting the explicit form of the potential. The chosen potentials have the benefit of permit an analytical approach [7,8] as well as leads to the solutions by means of topological defects.

\section{Model I: polynomial superpotential}

In this section we consider a model described by the following superpotential

\begin{equation}
W(\phi_1,...,\phi_n) = \sum_{i=1}^n \lambda_i \bigg(\phi_i - \frac{\phi_i^3}{3}\bigg) = \lambda_1 \bigg(\phi_1 - \frac{\phi_1^3}{3}\bigg) + \lambda_2 \bigg(\phi_2 - \frac{\phi_2^3}{3}\bigg) + ... + \lambda_n \bigg(\phi_n - \frac{\phi_n^3}{3}\bigg) ,
\end{equation} 
that is, we are considering that each superpotential $W_i(\phi_i)$ is given by

\begin{equation}  \label{Wi}
W_i(\phi_i) = \lambda_i \bigg(\phi_i - \frac{\phi_i^3}{3}\bigg) .
\end{equation} 
In fact, the above superpotential may be understood as a generalization of the model considered in Ref. \cite{DEWOLFE}. Substituting the last expression in Eq. (\ref{firstorder2}) we obtain

\begin{equation}
\frac{d\phi_i}{dr} = \lambda_i(1 - \phi_i^2), \quad (i = 1,2,...,n).
\end{equation} 
The solutions of the above equation are the usual $\lambda \phi^4$ kink, which are given by

\begin{equation}  \label{kink}
\phi_i(r) = \tanh[\lambda_i(r - r_i)], \quad (i = 1,2,...,n),
\end{equation} 
where $r_i$ is an integration constant associated to the translational invariance, representing the center of the kink. As one can see each scalar field $\phi_i(r)$ represents a domain wall centered at $r_i$. Substituting Eq. (\ref{kink}) in (\ref{Wi}) and then substituting the result back into (\ref{Ai}), we obtain

\begin{equation}
\frac{dA_i}{dr} = -\frac{2}{3} \bigg( \tanh(\lambda_i \xi_i) - \frac{\tanh^3(\lambda_i \xi_i)}{3} \bigg),
\end{equation} 
where we use $\xi_i = r - r_i$. Integrating the above equation we obtain the following result

\begin{equation}
A_i(r) = C_i - \frac{1}{9} \bigg( \tanh^2(\lambda_i \xi_i) + 2 - 4 \ln[\sech(\lambda_i \xi_i)] \bigg),
\end{equation} 
being $C_i$ an integration constant. Summing over $i$, we get

\begin{equation}  \label{A(r)}
A(r) = C - \frac{1}{9} \left\{ 2n + \sum_{i=1}^n \tanh^2(\lambda_i \xi_i) - 4 \ln \prod_{i=1}^n \sech(\lambda_i \xi_i) \right\},
\end{equation} 
where $C = \sum_i C_i$.

In order to fix the integration constant $C$ we impose that $A(\tilde{r}) = 0 $, where $\tilde{r}$ is defined as the average value of the coordinates of center of the kinks

\begin{equation}
\tilde{r} =\frac{1}{n}\sum_{i=1}^n r_i.
\end{equation} 
Imposing the above condition we obtain

\begin{equation}
C = \frac{1}{9}\left\{2n + \sum_{i=1}^n \tanh^2(\lambda_i \tilde{\xi}_i) - 4 \ln \prod_{i=1}^n \sech(\lambda_i \tilde{\xi}_i) \right\},
\end{equation} 
where we denote $\tilde{\xi}_i = \tilde{r} - r_i$. Substituting this result in Eq. (\ref{A(r)}) we have

\begin{equation}
A(r) = \frac{1}{9}\sum_{i=1}^n \left[ \tanh^2(\lambda_i \tilde{\xi}_i) - \tanh^2(\lambda_i \xi_i) \right] + \frac{4}{9}\ln \bigg( \prod_{i=1}^n\frac{\sech(\lambda_i \tilde{\xi}_i)}{\sech(\lambda_i \xi_i)} \bigg).
\end{equation} Finally, the warp factor is given by

\begin{equation}  \label{warp1}
e^{2A(r)} = \bigg( \prod_{i=1}^n\frac{\sech(\lambda_i \tilde{\xi}_i)}{\sech(\lambda_i \xi_i)} \bigg)^{8/9} \times \prod_{i=1}^n \exp \bigg(\frac{2}{9}\left[\tanh^2(\lambda_i \tilde{\xi}_i) - \tanh^2(\lambda_i \xi_i) \right] \bigg).
\end{equation}

Having settled to model, we shall consider some particular cases, namely $n=2$ and $n=3$, in order to better analyze the results obtained in this section.

\subsection{Case $n=2$}

Taking $n=2$ in Eq. (\ref{warp1}), we may rewrite the warp factor as follows

\begin{eqnarray}
e^{2A(r)} &=& \left( \frac{\sech(\lambda_1 \tilde{\xi}_1) \sech(\lambda_2 \tilde{\xi}_2)}{\sech(\lambda_1 \xi_1)\sech(\lambda_2 \xi_2)} \right)^{8/9} \notag \\ &\times& \exp \left( \frac{2}{9}\left[ \tanh^2(\lambda_1 \tilde{\xi}_1) + \tanh^2(\lambda_2 \tilde{\xi}_2) - \tanh^2(\lambda_1 \xi_1) - \tanh^2(\lambda_2 \xi_2) \right]\right).
\end{eqnarray}

\begin{figure}[H]
\begin{center}
\includegraphics[width=8cm]{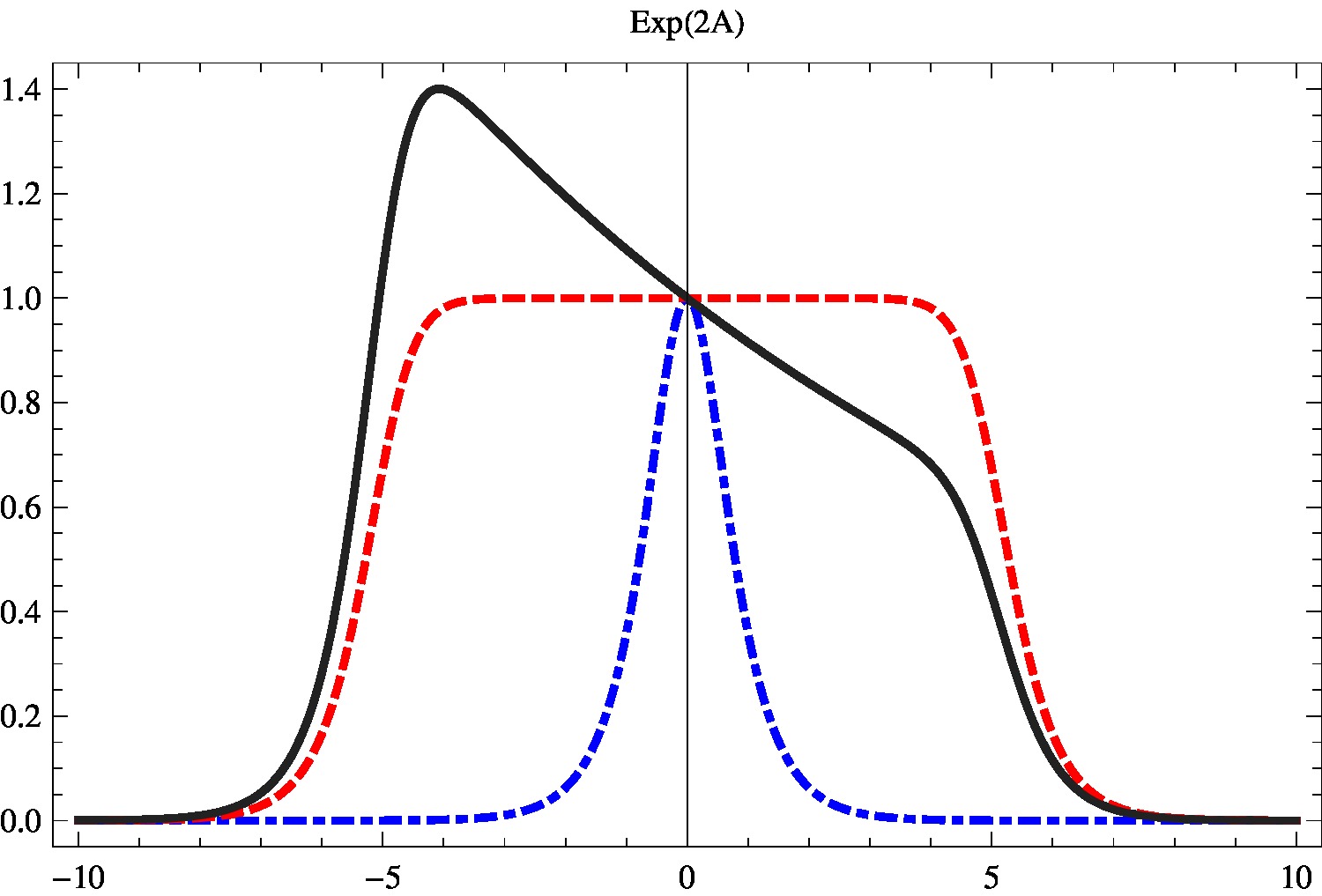}
\end{center}
\caption{\footnotesize {Warp factor - model I, case $n=2$. Where $r_1 = r_2 = 0$ and $\protect\lambda_1 = \protect\lambda_2 = 1$
(dotted-dashed line); $r_1 = -r_2 = 5$ and $\protect\lambda_1 = \protect\lambda_2 = 1$ (dashed line); $r_1 = -r_2 = 5$, $\protect\lambda_1 = 0.9$ and $\protect\lambda_2 = 1$ (solid line).}}
\label{fig1}
\end{figure}

In Fig. \ref{fig1} we plot the warp factor for three different values of the parameters $r_1$, $r_2$, $\lambda_1$ and $\lambda_2$. Note that $r_1$ and $r_2$ performs a kind of splitting brane mechanism, while $\lambda_1$ and $\lambda_2$ controls the symmetry of the problem. In the figures 2 and 3 we plot, respectively, the energy density and the scalar curvature. \newline

\begin{figure}[H]
\center
\includegraphics[width=8cm]{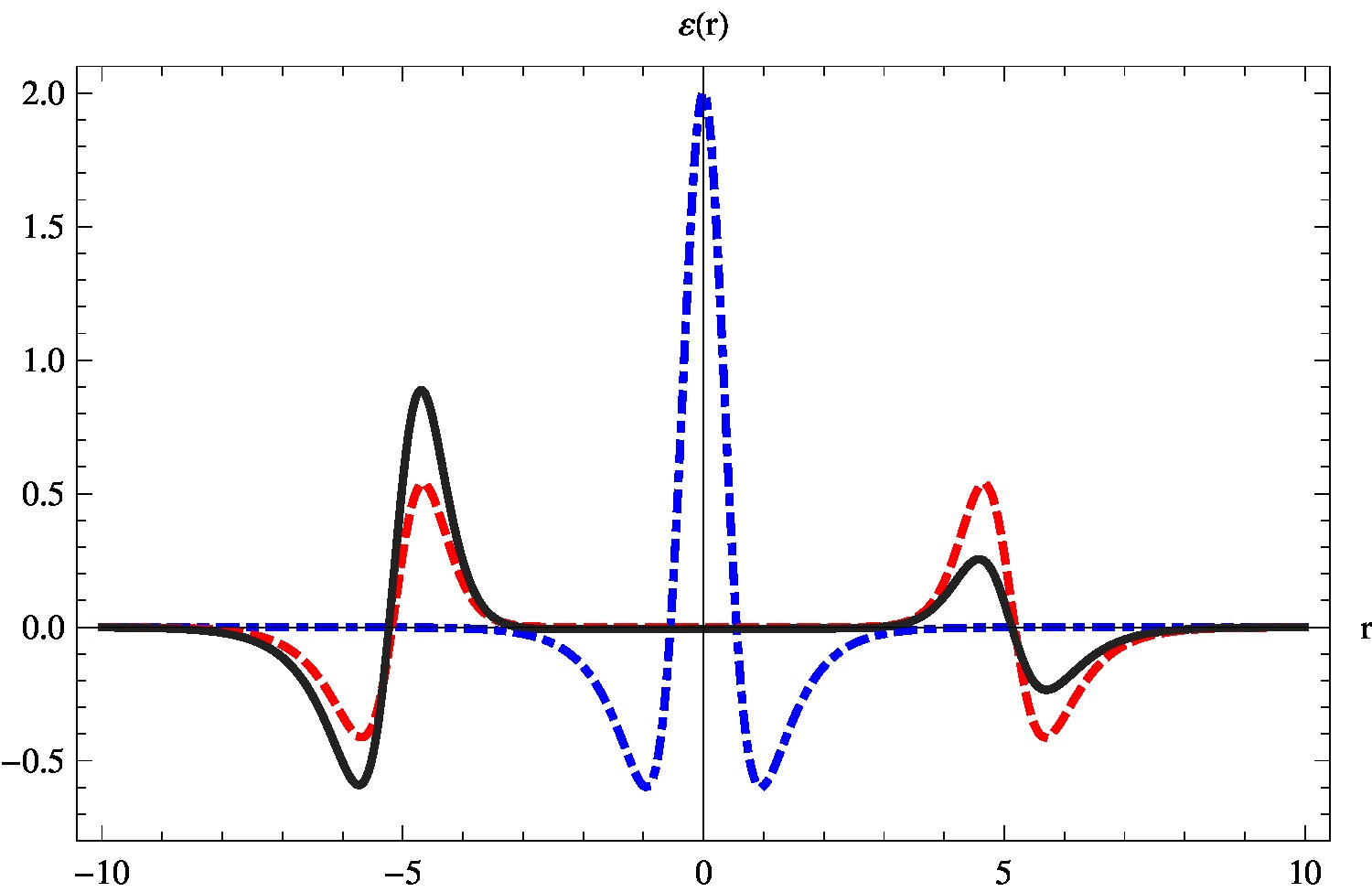}
\caption{\footnotesize {Energy density. The coventions (color and line codes) are the same as in Fig. \ref{fig1}.}}
\label{fig2}
\end{figure}

\begin{figure}[H]
\center
\includegraphics[width=8cm]{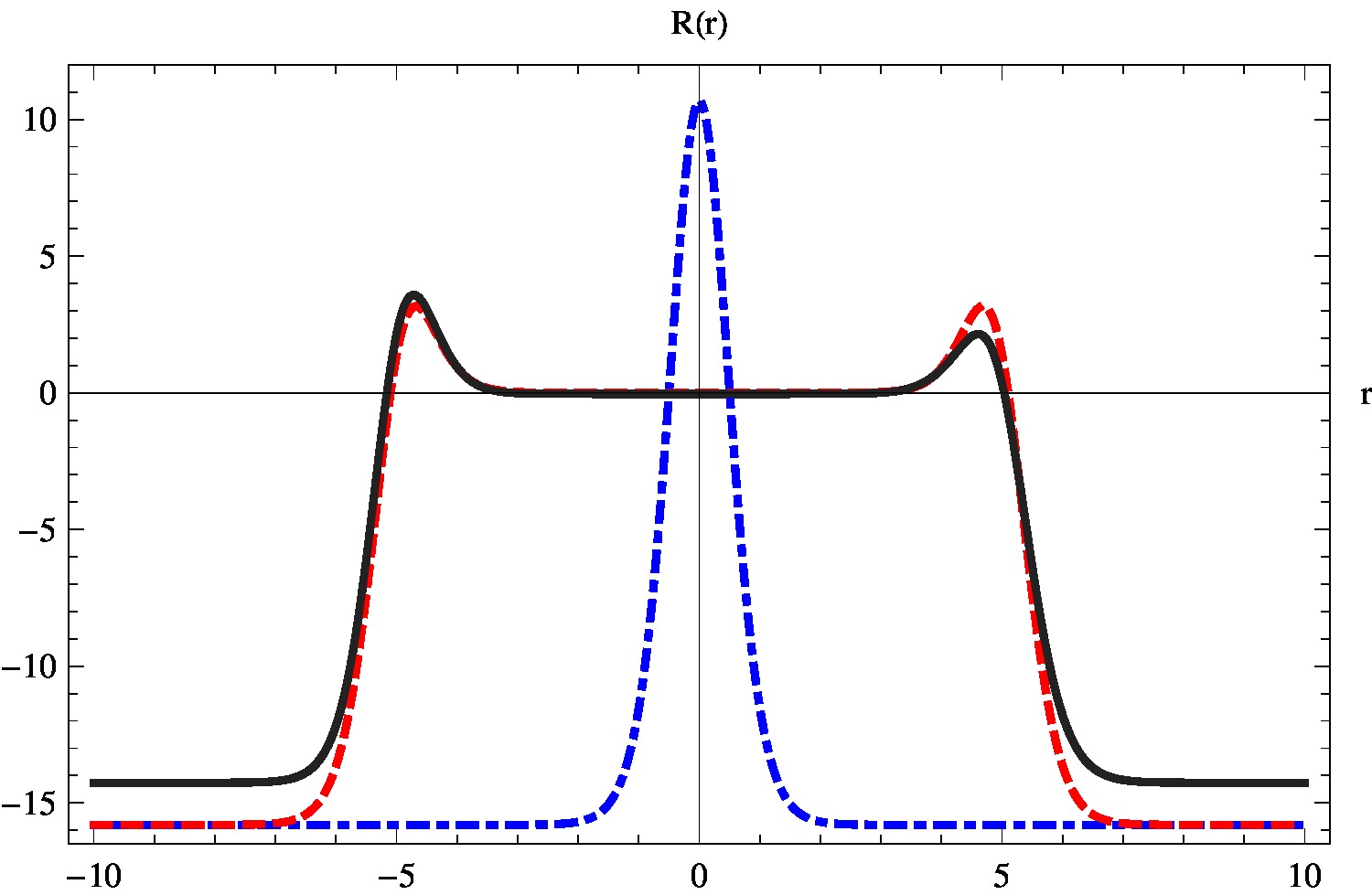} 
\caption{\footnotesize {Scalar of curvature. The coventions (color and line codes) are the same as in Fig. \ref{fig1}.}}
\label{fig3}
\end{figure}

The explicit dotted-dashed line in the above figures shows the existence of one brane engendering a quite symmetrical warp factor. The dashed line case also expresses a symmetrical warp factor, but now Figs. \ref{fig2} and \ref{fig3} demonstrate that the better interpretation shall be done in terms of two different branes whose core are positioned at $r_1$ and $r_2$. Finally, from Fig. \ref{fig1} it is possible to appreciate the possible use of these type of models to approach the hierarchy problem in the bottom case. In principle, every Higgs-like potential in an effective four-dimensional theory is sensible to an asymmetric warp factor. In this sense, it would be possible to compare the energy scales of the branes. In this vein, the value $e^{A(r_2) - A(r_1)} $ may be related to the energy gap between the branes, which, in turn, may be associated to the gap of fundamental and TeV scales.

\subsection{Case $n=3$}

\begin{figure}[H]
\center
\includegraphics[width=8cm]{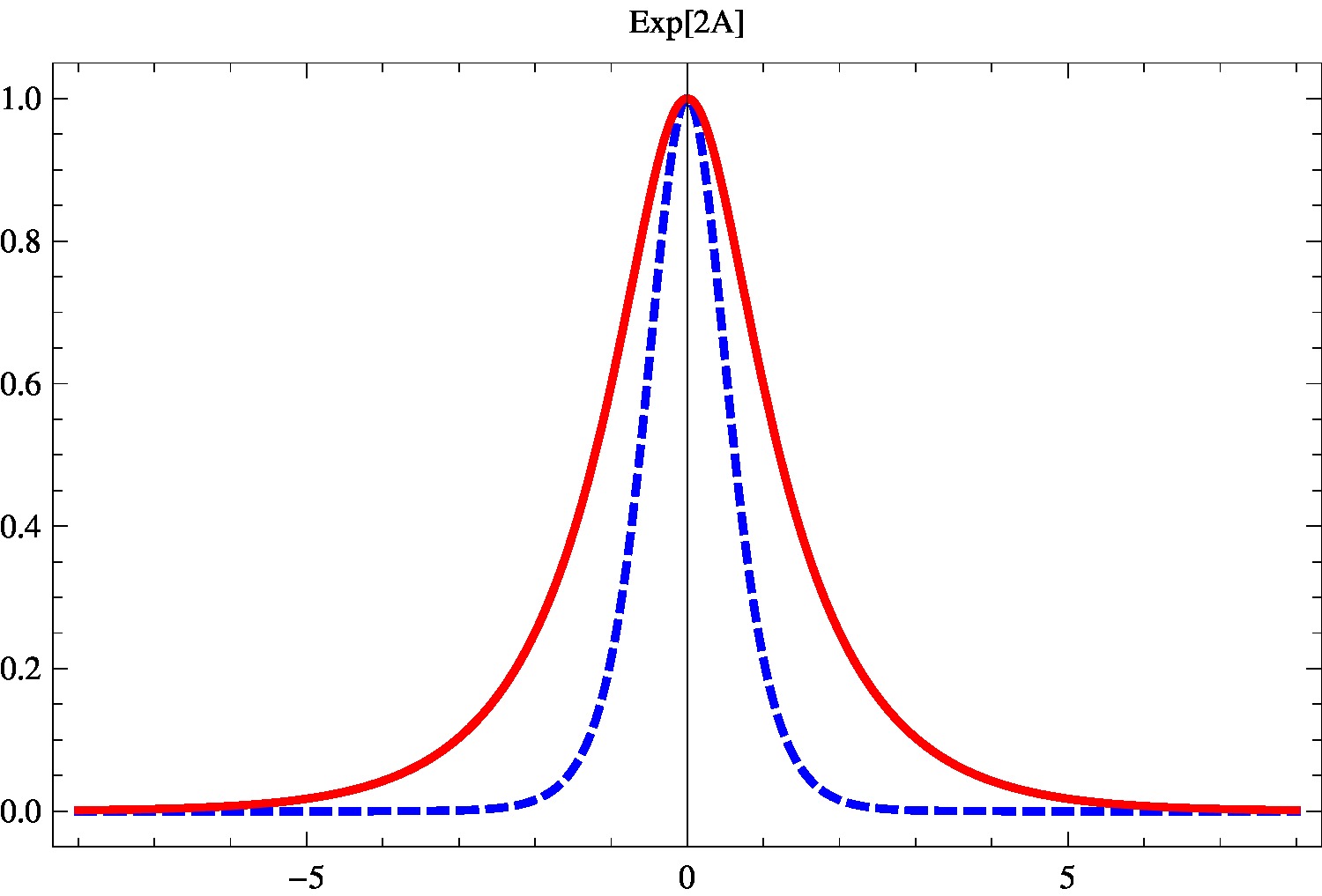}
\caption{{\protect\footnotesize {Warp factor - model I, case $n=3$. Where $r_1 = r_2 = r_3 = 0$ and $\protect\lambda_1 = \protect\lambda_2 = \protect\lambda_3 = 1$ (dashed line); $r_1 = -r_3 = 10$, $r_2 = 0$, $\protect\lambda_1 = \protect\lambda_ 2 = \protect\lambda_3 = 1$ (solid line).}}}
\label{fig4}
\end{figure}

\begin{figure}[H]
\center
\includegraphics[width=8cm]{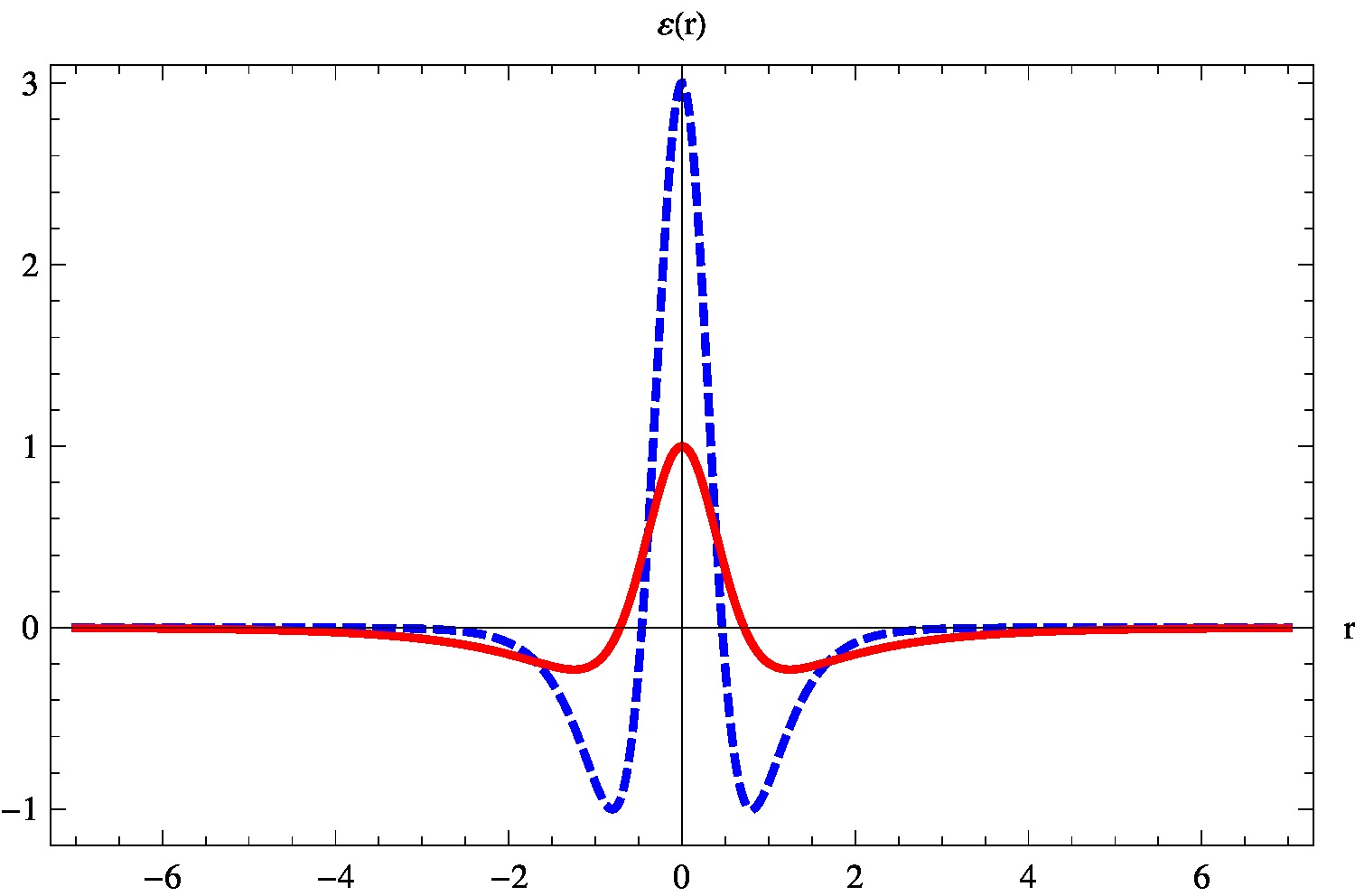}
\caption{\footnotesize {Energy density. The coventions (color and line codes) are the same as in Fig. \ref{fig4}.}}
\label{fig5}
\end{figure}

\begin{figure}[H]
\center
\includegraphics[width=8cm]{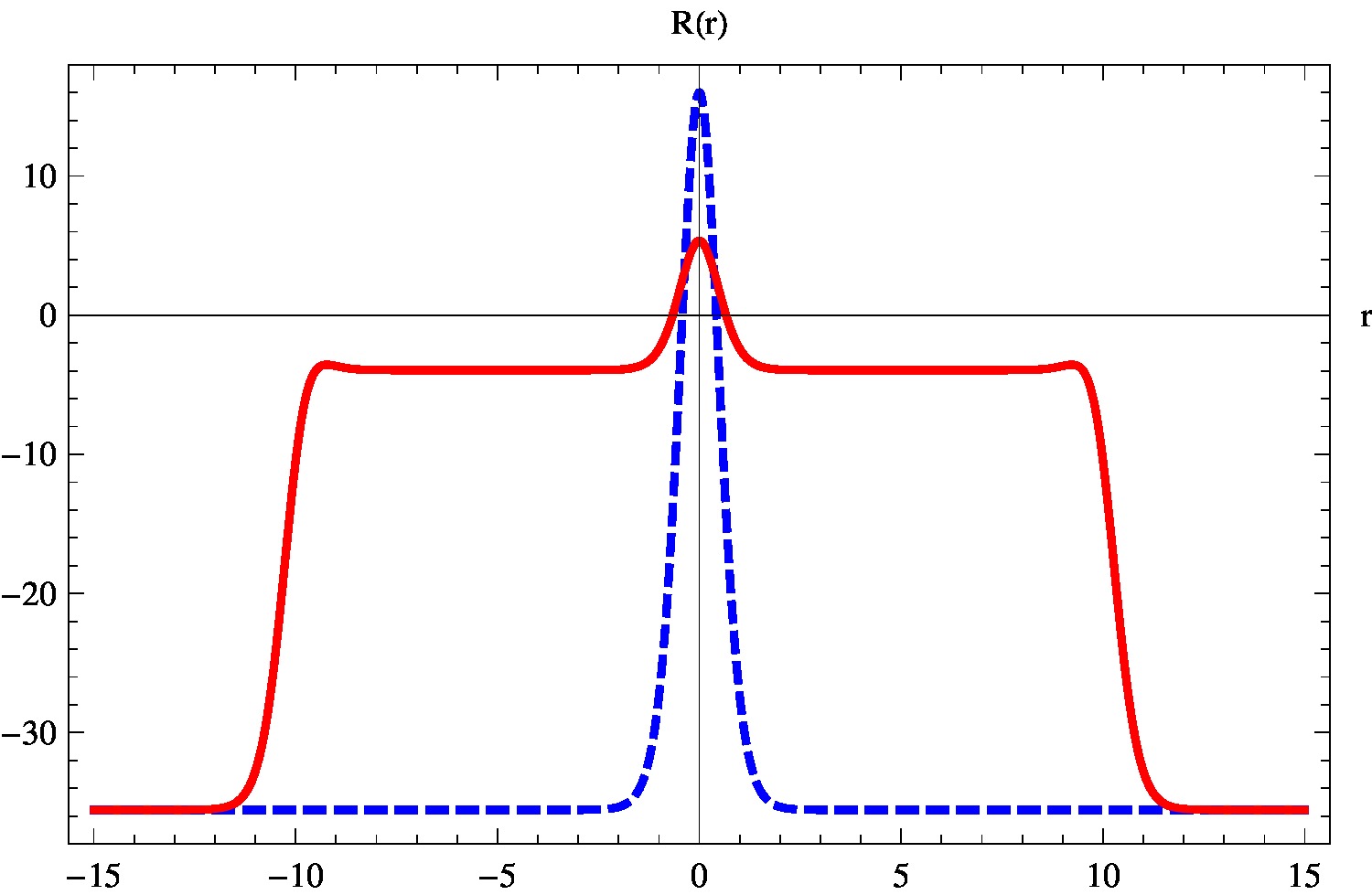}
\caption{\footnotesize{Scalar of curvature. The coventions (color and line codes) are the same as in Fig. \ref{fig4}.}}
\label{fig6}
\end{figure}

In this case, using $n=3$ in Eq. (\ref{warp1}), we may rewrite the warp factor as follows

\begin{eqnarray}
& e^{2A(r)}& = \(\frac{\sech(\lambda_1 \tilde{\xi}_1) \sech(\lambda_2 \tilde{\xi}_2)\sech(\lambda_3 \tilde{\xi}_3)}{\sech(\lambda_1 \xi_1)\sech(\lambda_2 \xi_2)\sech(\lambda_3 \xi_3)} \)^{8/9}  \nonumber\\ &\times& \exp\Bigg(\frac{2}{9}\left[ \tanh^2(\lambda_1 \tilde{\xi}_1) +  \tanh^2(\lambda_2 \tilde{\xi}_2) + \tanh^2(\lambda_3 \tilde{\xi}_3) \right]\Bigg) \nonumber\\ &\times & \exp\Bigg(-\frac{2}{9}\left[\tanh^2(\lambda_1 \xi_1) + \tanh^2(\lambda_2 \xi_2) + \tanh^2(\lambda_3 \xi_3) \right]\Bigg).
\end{eqnarray} 
In Fig. \ref{fig4} it is shown the plot of the above warp factor for two cases. The respective energy densities are shown in Fig. \ref{fig5}. However, Figs. \ref{fig4} and \ref{fig5} are not particularly useful to see the background properly. In Fig. \ref{fig6} we plotted scalar of curvature. The solid line in Fig. \ref{fig6} supports the interpretation of the formation of three branes, corresponding to those transition regions in the scalar of curvature, whose core are positioned at $r_1$, $r_2$ and $r_3$. We
see, from the warp factor shape, that in this case it is not viable to approach the hierarchy problem. It still interesting, however, to see how a plethora of braneworld models can be generated from the general formulation of Section II.

It is worth to emphasize that the energy density shape is constrained by the warp factor behavior in a general way (see, for instance, Eq. (\ref{energy})). Qualitatively, the net result of the warp factor can be understood as the product of the warp factors constituting the model in question. Since each warp factor is concentrated around the correspond domain wall, it is straightforward to see that the overlap warp factor region will be the most pronounced one for the effective result. As far as the domain walls are symmetrically 
distributed around the zero point (in the extra coordinate) the peak over this point is indeed expected.


\section{Model II: extension of Gremm's model}

The second model considered in this paper is described by the following superpotential

\begin{equation}
W(\phi_1,...,\phi_n) = \sum_{i=1}^n \lambda_i \sin \phi_i = \lambda_1 \sin \phi_1 + \lambda_2 \sin \phi_2 + ... + \lambda_n \sin \phi_n ,
\end{equation} 
that is, each superpotential $W_i(\phi_i)$ is given by

\begin{equation}  \label{Wi2}
W_i(\phi_i) = \lambda_i \sin \phi_i .
\end{equation} 
As one can see, this superpotential may be understood as an extension of the model considered by Gremm in Ref. \cite{GREMM}. The first-order equations for the scalar fields reads

\begin{equation}
\frac{d \phi_i}{d r} = \lambda_i \cos \phi_i, \quad (i = 1,2,...,n),
\end{equation} 
whilst the solutions of the above equations are the well known sine-Gordon kinks, which are given by

\begin{equation}
\phi_i(r) = \arctan(\sinh(\lambda_i \xi_i)), \quad \textmd{where}\quad \xi_i = r - r_i.
\end{equation} 
Substituting the above expression in Eq. (\ref{Wi2}) and then substituting it back in Eq. (\ref{Ai}) we obtain, after integration,

\begin{equation}
A_i(r) = C_i + \frac{2}{3}\ln \sech(\lambda_i \xi_i),
\end{equation} 
where $C_i$ is an integrating constant. Summing over $i$, we obtain the following expression

\begin{equation}  \label{A(r)2}
A(r) = C + \frac{2}{3}\ln \bigg(\prod_{i=1}^n \sech(\lambda_i \xi_i)\bigg).
\end{equation} 
By imposing $A(\tilde{r}) = 0$, where $\tilde{r} = \sum_{i=1}^n r_i/n$, we conclude that

\begin{equation}
C = - \frac{2}{3}\ln \bigg(\prod_{i=1}^n \sech(\lambda_i \tilde{\xi}_i)\bigg), \quad \textmd{where}\quad \tilde{\xi}_i = \tilde{r} - r_i.
\end{equation} Substituting the last equation in Eq. (\ref{A(r)2}), we arrive at

\begin{equation}
A(r) = \frac{2}{3}\ln \bigg(\prod_{i=1}^n \frac{\sech(\lambda_i \xi_i)}{\sech(\lambda_i \tilde{\xi}_i)}\bigg).
\end{equation} Finally, the warp factor obtained for this model is given by the following
expression

\begin{equation}
e^{2A(r)} = \bigg(\prod_{i=1}^n \frac{\sech(\lambda_i \xi_i)}{\sech(\lambda_i \tilde{\xi}_i)}\bigg)^{4/3}.
\end{equation}

\subsection{Case $n=2$}

In the particular case where $n=2$ we obtain the following warp factor

\begin{equation}
e^{2A(r)} = \bigg(\frac{\sech(\lambda_1 (r - r_1)) \sech(\lambda_2 (r - r_2))}{ \sech(\lambda_1 (\tilde{r} - r_1)) \sech(\lambda_2 (\tilde{r} - r_2))}\bigg)^{4/3}.
\end{equation}

\begin{figure}[h]
\center
\includegraphics[width=8cm]{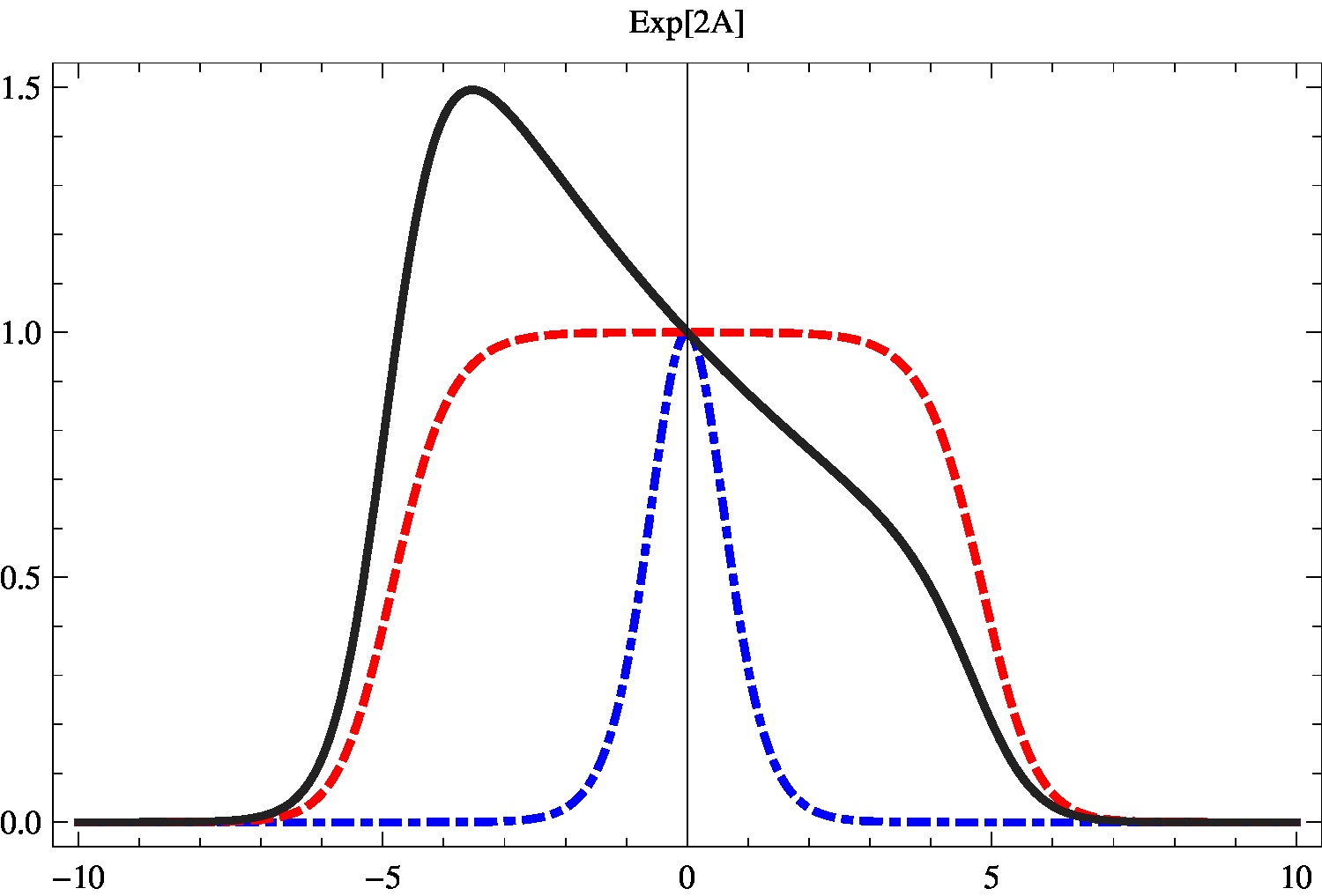} 
\caption{\footnotesize {Warp factor - model II, case $n=2$. Where $r_1 = r_2 = 0$ and $\protect\lambda_1 = \protect\lambda_2 = 1$
(dotted-dashed line); $r_1 = -r_2 = 5$ and $\protect\lambda_1 = \protect \lambda_2 = 1$ (dashed line); $r_1 = -r_2 = 5$, $\protect\lambda_1 = 0.9$ and $\protect\lambda_2 = 1$ (solid line).}}
\label{fig7}
\end{figure}

\begin{figure}[H]
\center
\includegraphics[width=8cm]{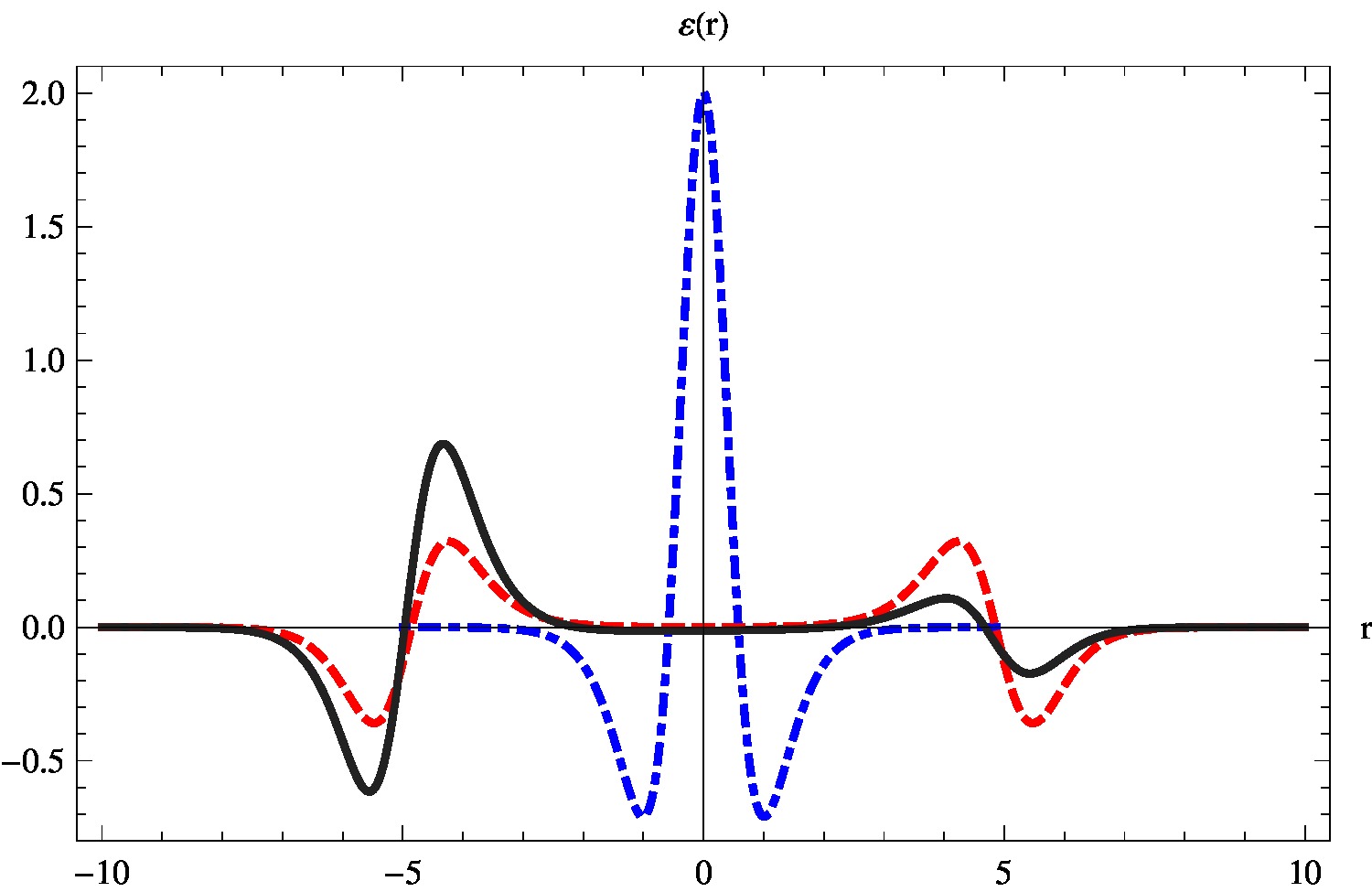} 
\caption{\footnotesize {Energy density. The coventions (color and line codes) are the same as in Fig. \ref{fig7}.}}
\label{fig8}
\end{figure}

\begin{figure}[H]
\center
\includegraphics[width=8cm]{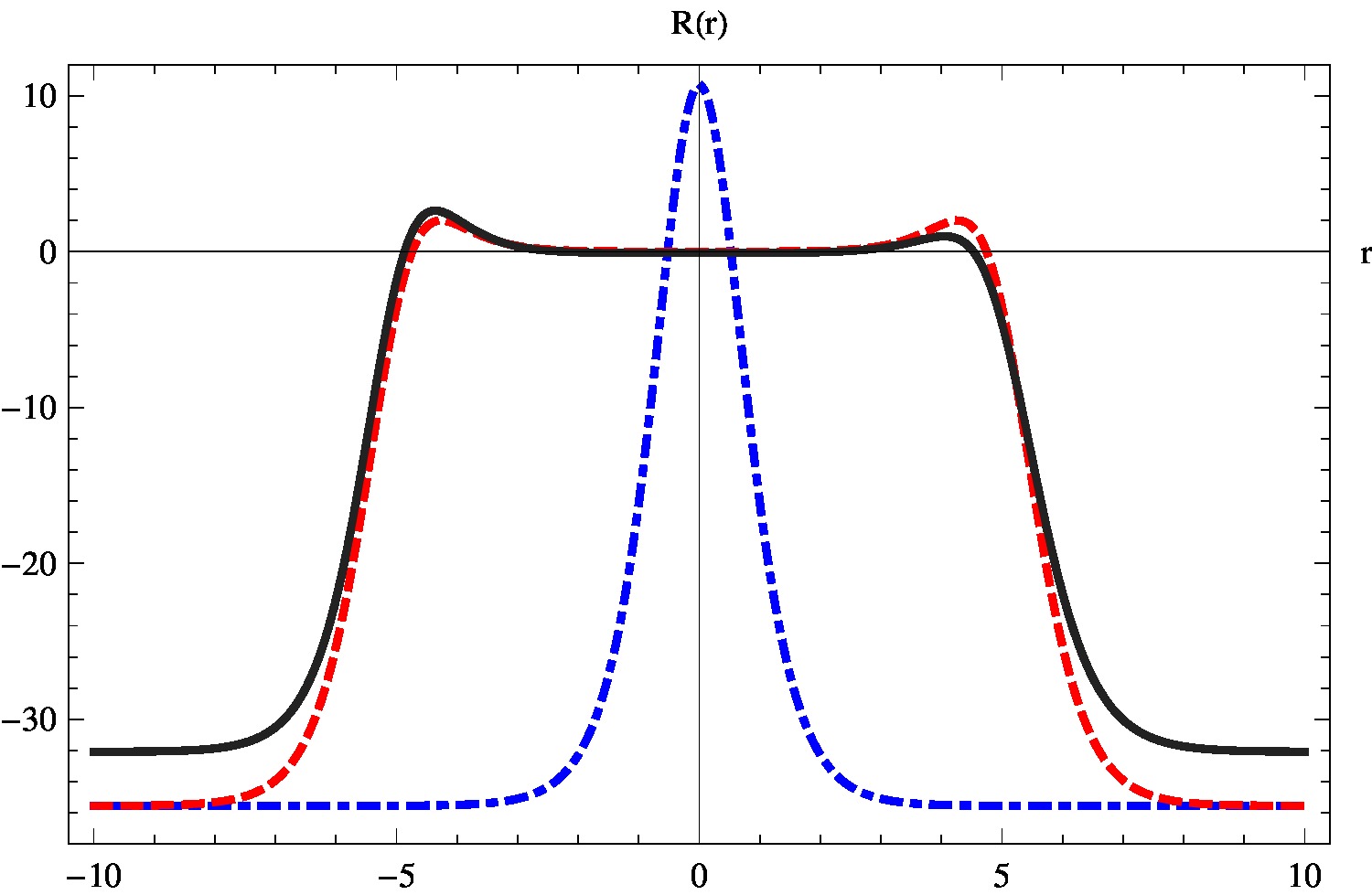} 
\caption{\footnotesize{Scalar of curvature. The coventions (color and line codes) are the same as in Fig. \ref{fig7}.}}
\label{fig9}
\end{figure}
As one can see in Figs. \ref{fig7}, \ref{fig8} and \ref{fig9}, the results obtained for this model in the case $n=2$ are very similar to the results obtained for model I (also for $n=2$) and, therefore, the same qualitative discussion performed to that case may be applied here. Regarding these common features, it can be understood from the fact that for a given topological sector, the potentials are quite similar. As a consequence, the according solutions and behaviors share many properties.

\subsection{Case $n=3$}

Using $n=3$, it is straightforward to see that we may rewrite the warp factor as

\begin{equation}
e^{2A(r)} = \bigg(\frac{\sech(\lambda_1 (r - r_1)) \sech(\lambda_2 (r - r_2)) \sech (\lambda_3 (r - r_3))}{\sech(\lambda_1 (\tilde{r} - r_1)) \sech(\lambda_2 (\tilde{r} - r_2)) \sech(\lambda_3 (\tilde{r} - r_3))}\bigg)^{4/3}.
\end{equation}

\begin{figure}[H]
\center
\includegraphics[width=8cm]{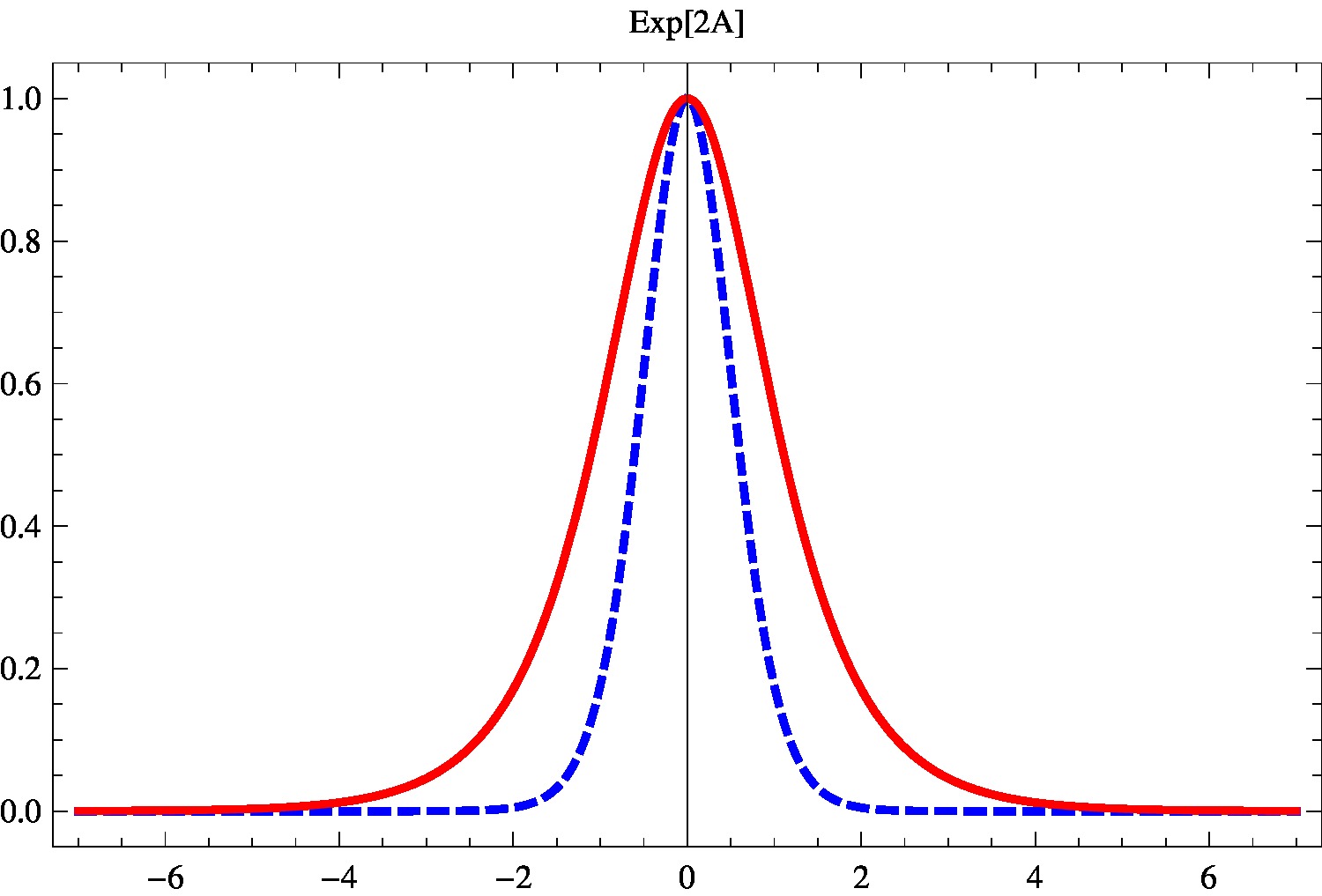} 
\caption{\footnotesize{Warp factor - model II, case $n=3$. Where $r_1 = r_2 = r_3 = 0$ and $\protect\lambda_1 = \protect\lambda_2 = \protect \lambda_3 = 1$ (dashed line); $r_1 = -r_3 = 10$, $r_2 = 0$, $\protect\lambda_1 = \protect\lambda_ 2 = \protect\lambda_3 = 1$ (solid line).}}
\label{fig10}
\end{figure}

\begin{figure}[H]
\center
\includegraphics[width=8cm]{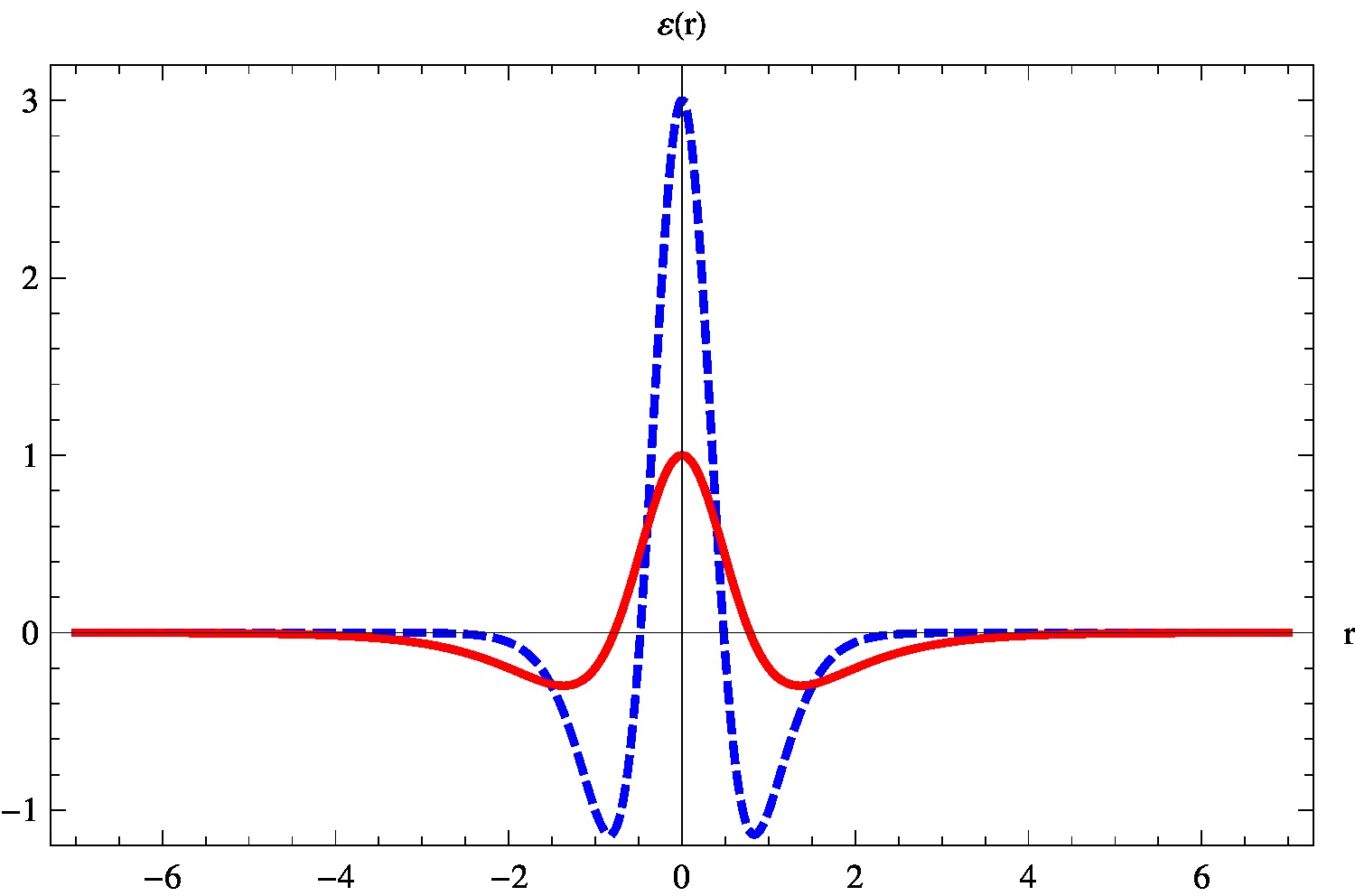} 
\caption{\footnotesize{Energy density. The coventions (color and line codes) are the same as in Fig. \ref{fig10}.}}
\label{fig11}
\end{figure}

\begin{figure}[H]
\center
\includegraphics[width=8cm]{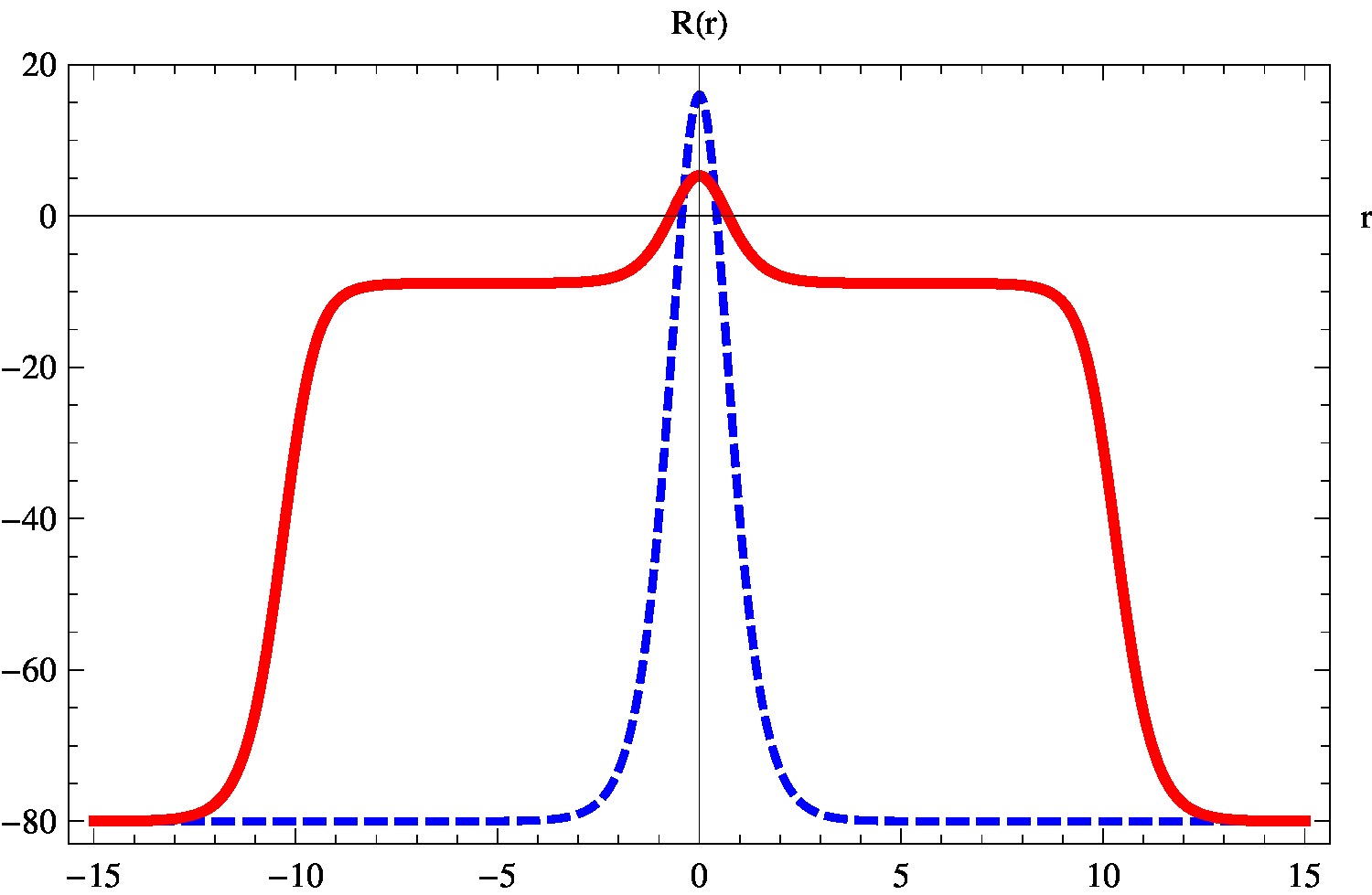} 
\caption{\footnotesize {Scalar of curvature. The coventions (color and line codes) are the same as in Fig. \ref{fig10}.}}
\label{fig12}
\end{figure}

Once again, as one can see in Figs. \ref{fig10}, \ref{fig11} and \ref{fig12}, the graphical results show a huge resemblance with model I (for $n=3$), the differences being at the detail level. Hence, the same qualitative behavior is expected here.

\section{STABILITY AND METRIC FLUCTUATION}

In this section we investigate the metric fluctuation regarding the braneworld scenario. For the metric fluctuations we adopt a gauge where the perturbed interval becomes \cite{GREMM,DEWOLFE}

\begin{equation}
ds^2 = e^{2A(r)}(\eta_{\mu\nu} + h_{\mu\nu})dx^{\mu}dx^{\nu} + dr^2,
\end{equation} 
where $h_{\mu\nu} = h_{\mu\nu}(x^{\alpha},r)$ represents small perturbations. In general, we may also consider small perturbations around
classical solutions of the scalar fields, but in this case, the set of differential equations obtained are too complicated. However, the situation may be simplified if one consider only the transverse and traceless sector of the metric fluctuation, which may be obtained by acting with the projector operator in $h^{\alpha \beta}$, that is $\bar{h}_{\mu\nu} = P_{\mu\nu\alpha\beta}h^{\alpha \beta}$ (for further details see \cite{DEWOLFE}). In this sector the metric fluctuation decouples from the scalar fields and we obtain the following equation

\begin{equation}
\frac{\partial^{2}\bar{h}_{\mu \nu }}{\partial r^{2}}+4\,\frac{dA}{dr}\, \frac{\partial\bar{h}_{\mu\nu }}{\partial r}=e^{-2\,A}\partial _{\rho}\partial ^{\rho}\bar{h}_{\mu \nu }.
\end{equation}
Now, performing the function redefinition $\bar{h}_{\mu \nu }(x^{\alpha},r) = e^{i\vec{k}.\vec{x}}e^{-\frac{3}{2}A(r)} \psi_{\mu \nu }(r)$ and the variable transformation $z = \int e^{-A(r)} dr$, we can recast the above equation as a quantum mechanics-like problem

\begin{equation}
-\frac{d^2 \psi_{\mu\nu}}{dz^2} + U_{eff}(z)\psi_{\mu\nu} = k^2 \psi_{\mu\nu},
\end{equation} 
where the effective potential is defined by

\begin{equation}
U_{eff}(z) = \frac{9}{4}\left(\frac{dA}{dz}\right)^2 + \frac{3}{2}\frac{d^2 A}{dz^2}.
\end{equation} 
In terms of the variable $r$ the effective potential may be written as

\begin{equation}
U_{eff}(r) = \frac{3}{4}e^{2A(r)} \bigg( 5 \left(\frac{dA}{dr}\right)^2 + 2 \frac{d^2 A}{dr^2} \bigg).
\end{equation} 
In order to analyze the stability we note that the Hamiltonian-like operator, defined by $\hat{H} = -\frac{d^2 }{dz^2} + U_{eff}(z)$, can be factorized as $\hat{H} = \hat{S}^{\dagger}\hat{S}$, where

\begin{equation}
\hat{S} = \frac{d}{dz} - \frac{3}{2}\frac{dA}{dz} \quad \textmd{and} \quad \hat{S}^{\dagger} = - \frac{d}{dz} - \frac{3}{2}\frac{dA}{dz},
\end{equation} 
and consequently the eigenvalues $k^2$ are strictly positives, which ensures the stability with respect to the metric fluctuations. The zero mode $\psi^{(0)}_{\mu\nu}$ ($k^2 = 0$) may be easily obtained by noticing that the action of operator $\hat{S}$ on $\psi_0$ should vanish, that is

\begin{equation}
\hat{S}\psi^{(0)}_{\mu\nu} = \bigg(\frac{d}{dz} - \frac{3}{2}\frac{dA}{dz} \bigg) \psi^{(0)}_{\mu\nu} = 0 .
\end{equation}

Integrating the above equation with respect to $z$, we obtain

\begin{equation}
\psi^{(0)}_{\mu\nu}(z) = N_{\mu\nu} e^{\frac{3}{2}A(z)},
\end{equation}

\noindent where $N_{\mu\nu}$ is a normalization factor. Returning, then, to the variable $r$ we get simply

\begin{equation}
\psi^{(0)}_{\mu\nu}(r) = N_{\mu\nu} e^{\frac{3}{2}A(r)} = N_{\mu\nu} \prod_{i=1}^n e^{\frac{3}{2}A_i(r)}.
\end{equation}

After investigating the tensorial modes, it is also necessary to approach the vectorial and scalar modes. In fact, in order to fully guarantee stability these perturbations must be addressed. We emphasize that most of the necessary analysis concerning warped braneworlds can be found in the Ref. \cite{ahmed}. Here we shall make strong use of the study performed in \cite{ahmed}, stressing the most relevant points which are particular to our case. 

Taking into account vectorial perturbations, nothing but considering the decomposition $\partial_\mu G_\nu+\partial_\nu G_\mu$ as a possible perturbative mode and working with the transverse sector, it is possible to see that the canonical normal modes cannot be localized on the branes (see Ref. \cite{ahmed}, Section 6.3). Actually, in the absence of an {\it ad hoc} mechanism it is not possible to localize vector fields on a given brane. This is because any rank-two term appearing in the five dimensional action has zero gravitational weight (see, for instance, Ref. \cite{chumbes} for a discussion in a broader context). Therefore vectorial modes do not have any effect in stability issue concerning warped braneworlds.

The reasoning used in the previous paragraph cannot be used in any way for scalar modes. Taking into account scalar perturbations to the metric in the longitudinal gauge we have 
\begin{eqnarray}
ds^2=e^{2A}(1-2\psi)\eta_{\mu\nu}dx^\mu dx^\nu+(1+2\chi)dr^2.\label{new1}
\end{eqnarray} It turns out that the linearized field equations to the perturbation gives as a first outcome (see, again, Ref. \cite{ahmed} for all the details) the constraint $\chi=2\psi$. In the so-called conformal frame, given by the transformation $dr=e^{A(r)}dz$, these equations reads
\begin{eqnarray}
\ddot{\psi}+\dot{A}\dot{\psi}-\Box\psi=\frac{4}{3}\sum_{i=1}^{n}\dot{\phi}_i\dot{\varphi}_i, \label{new2}\\
2\dot{A}\psi+\dot{\psi}=\frac{2}{3}\sum_{i=1}^{n}\dot{\phi}_i\varphi_i,\label{new3}
\end{eqnarray} 
where a dot means derivative with respect to the $z$ coordinate and $\varphi_i$ is the corresponding perturbation of the scalar field $\phi_{i}$. Let us write the scalar fields perturbations as $\varphi_i = f_i(x^\mu,z)+\varphi(x^\mu,z)$, in such a way that a part of the perturbations is common to all scalar fields. Notice that this type of perturbation split does not preclude the individuality of each perturbation, encoded this time in $f_i(x^\mu,z)$. Substituting this split into Eqs. (\ref{new2}) and (\ref{new3}), differentiating Eq. (\ref{new3}) and using (\ref{new2}) we have
\begin{eqnarray}
\ddot{\psi}+3\dot{A}\dot{\psi}+4\ddot{A}\psi+\Box\psi=\frac{4}{3}\Bigg[\sum_{i=1}^{n}\ddot{\phi}_if_i+\varphi\sum_{i=1}^n\ddot{\phi}_i\Bigg].\label{new4}
\end{eqnarray} By isolating the $\varphi$ term with the aid of Eq. (\ref{new3}), we may recast Eq. (\ref{new4}) as
\begin{eqnarray}
\ddot{\psi}+\Bigg[3\dot{A}-2\frac{(\sum_{i=1}^n\ddot{\phi}_i)}{(\sum_{i=1}^n\dot{\phi}_i)}\Bigg]\dot{\psi}+\Bigg[4\ddot{A}-4\dot{A}\frac{(\sum_{i=1}^n\ddot{\phi}_i)}{(\sum_{i=1}^n\dot{\phi}_i)}+\Box\Bigg]\psi+ \sum_{i=1}^n g_i(z) f_i =0,\label{new5}
\end{eqnarray} 
where $g_i(z)$ is given by
\begin{eqnarray}
g_i(z) = \frac{4}{3} \Bigg[ \frac{(\sum_{j=1}^n \ddot{\phi}_j)}{(\sum_{k=1}^n \ddot{\phi}_k)} \dot{\phi}_i -  \ddot{\phi}_i \Bigg] \label{new6}
\end{eqnarray} 
Henceforward the procedure is quite similar to the usual study of scalar perturbation. Considering the following redefinitions
\begin{equation}
\psi = e^{-3 A(z)/2}\tilde{\psi}\sum_{i=1}^n\dot{\phi}_i,\label{new7}
 \quad \textmd{and} \quad
 g_i = e^{-3 A(z)/2}\tilde{g_i}\sum_{i=1}^n\dot{\phi}_i ,
\end{equation}
it is possible to rewrite Eq. (\ref{new6}) in a Schrödinger-like form, namely
\begin{equation}
- \ddot{\tilde{\psi}} + U_{eff}(z) \tilde{\psi} = \Box \tilde{\psi} + \sum_{i=1}^n \tilde{g}_i f_i ,
\end{equation}
where the effective potential is
\begin{eqnarray}
U_{eff}=\frac{9}{4}\dot{A}^2-\frac{5}{2}\ddot{A}+\dot{A}\frac{(\sum_{i=1}^n\ddot{\phi}_i)}{(\sum_{i=1}^n\dot{\phi}_i)}+2\Bigg(\frac{(\sum_{i=1}^n\ddot{\phi}_i)}{(\sum_{i=1}^n\dot{\phi}_i)}\Bigg)^2-\frac{(\sum_{i=1}^n\dddot{\phi}_i)}{(\sum_{i=1}^n\dot{\phi}_i)}.\label{new9}
\end{eqnarray}

By using the standard decomposition $\tilde{\psi}= \int d^4p \, e^{ip_\mu x^\mu} \bar{\psi}_p(z)$, so that $\Box \tilde{\psi} = m^2 \tilde{\psi}$ (where $p_\mu p^\mu = - m^2$). We also consider the following representation $\tilde{f}_i = \int d^4p \, e^{ip_\mu x^\mu} \bar{f}_{i_p}(z)$. In this vein, the four-dimensional mass of the fluctuation in also taken into account. The resulting equation reads 
\begin{eqnarray}
-\ddot{\bar{\psi}}_p + U_{eff}\bar{\psi}_p=m^2\bar{\psi}_p +  \sum_{i=1}^n \tilde{g}_i \bar{f}_{i_p}(z). \label{new8}
\end{eqnarray}
Surprisingly enough it is possible to rewrite Eq. (\ref{new8}) in a suitable operatorial form by using the definition 
\begin{equation}
\alpha=\frac{e^{3A/2}\sum_{i=1}^n\dot{\phi}_i}{\dot{A}}. \label{new10}  
\end{equation} In fact, it can be verified that Eq. (\ref{new8}) can be written as
\begin{equation}
\mathbb{A}^\dagger\mathbb{A}\bar{\psi}=m^2\bar{\psi}+\sum_{i=1}^n \tilde{g}_i \bar{f}_{i_p}(z),\label{new11}
\end{equation}being $\mathbb{A}=\partial(z)+\omega$ ($\mathbb{A}^\dagger=-\partial(z)+\omega$), with $\omega=\frac{\dot{\alpha}}{\alpha}$. The absence of taquionic modes is evinced from the homogeneous part of $\bar{\psi}$, since the particular sector has no influence in the mass spectrum.

\section{Concluding Remarks}

In this work we introduced a general method that can be used to construct analytical braneworld models coming from an arbitrary number of scalar fields. Interestingly, the models obtained through the use of this method are direct compositions of one-field models in the usual Minkowski space-time but, in the braneworld scenario a very nontrivial result shows up. After a general introduction of the approach, we implemented the idea in two situations, the first one where polynomial superpotentials are used in the described construction, and a second one using periodic sine-Gordon type superpotentials were used.

It is important to remark that the method is not restricted to the use of the combination of one-field nonlinear superpotentials. In fact one could combine two or more superpotentials where two or more coupled scalar fields appear non trivially, as the ones used in a number of publications \cite{gomes,amaro,hoff}.

The braneworld scenarios established here present the interesting features of brane splitting and, in some cases, an asymmetric warp factor. The type of extension performed in obtaining these models brings a twofold characteristic in its scope: on the one hand it is a slight mathematical extension of the usual approach, not bringing much additional technical difficulty. On the other hand, as shown, it may generate relevant results in the thick braneworld modeling.  

\section*{Acknowledgments}

ASD and JMHS are grateful to CNPq for financial support. GPB thanks to Fapesp for the financial support.

\end{document}